\documentclass[doublecol]{revtex4} 
\usepackage{graphicx}
\usepackage{amssymb,amsfonts}
\usepackage[latin1]{inputenc}
\usepackage{epstopdf}
\usepackage{xr}
\begin{document}

\title{Ray Chaos in a Photonic Crystal}

\author{Emmanuel Rousseau Didier Felbacq}

\affiliation{                    
  Laboratoire Charles Coulomb (L2C), UMR 5221 CNRS-Universit\'e de Montpellier, Montpellier, F-France\\
}

\begin{abstract}
The ray dynamics in a photonic crystal was investigated. Chaos occurs for perfectly periodic crystals, the rays dynamics being very sensitive to the initial conditions. Depending on the filling factor, the ray dynamics can exhibit stable paths near (fully) chaotic motion. The degree of chaoticity is quantified through the computation of Lyapunov exponents. As a result, the more diluted is the geometry, the more chaotic is the dynamic. Therefore, despite the perfect periodicity of the geometry, light transport is a diffusive process which can be tuned from normal diffusion (brownian motion) to anomalous diffusion because of the existence of L\'evy flights.
\end{abstract}

\maketitle

\section{Introduction and issue}

Photonic crystals, i.e. artificial periodic structures exhibiting a photonic band gaps \cite{Busch}, have been studied at very different angles since the pioneering works of S. John and E. Yablonovitch \cite{Sajeev,Yablo}. From the basic effect of acting like mirrors, the engineering of band structures and the recognition that they behave as metamaterials \cite{mesomag},  they have shown a wide range of physical phenomena. However, they have been hardly considered in the high frequency domain: it is indeed at most, say, the first three bands which are considered for band structure engineering or effective properties. \\
In this work, we consider the propagation of light in periodic dielectric structures when the wavelength is very small with respect to the scatterers size. In this limit, band diagrams become highly complicated because of a huge number of eigenmodes. This is the domain of wave chaos where puzzling phenomena such as random lasing \cite{Cao} or localization \cite{Albada,Wolf} may be expected.
In electromagnetism and optics, most of the studies dealt with chaotic cavities  \cite{Cao, Stockmann2, Hentschel, Aiello,Dingjan,Sweet} but only few of them with open chaotic systems such as optical fibers \cite{Doya} or photonic crystals\cite{Gumen,Cruz}. In the case of a two dimensional photonic crystal, chaos is revealed by the help of the statistical properties of the energy level spacing \cite{Gumen,Cruz}. They are computed from the eigenfrequencies of the Helmholtz equation solved with proper boundary conditions. These approaches might indicate that light dynamics in a 2D photonics-crystal is ``quasi-integrable" \cite{Cruz} meaning that the light dynamics exhibits regular and chaotic trajectories characterized through different level-spacing distributions. However, the dynamical properties of light are not a direct output of such an approach. Here we show that the dynamics of light and the transition from regular to chaotic trajectories can be understood with the help of light rays. We work then in the geometric approximation, the light rays representing the trajectories of the energy \cite{Doya}. In order to be valid, our model requires that the wavelength of light be at least ten times smaller than the cylinder radius (see section I of the Supplementary Materials for a detailed discussion). Within such an approximation,  we show that the transition from a regular motion to a chaotic one is imprinted in the dynamics of the transmitted rays, the dynamics of the reflecting rays  being always chaotic. Within the rays approximation, we compute the Lyapunov exponent that quantify the exponential sensitivity to the initial conditions. From the knowledge of the dynamics, we derive the diffusive properties of light in the photonic crystal. We show a transition from a brownian diffusion to an ``anomalous" regime. 

\begin{figure}[h]
\begin{center}
\includegraphics[width=4.5cm]{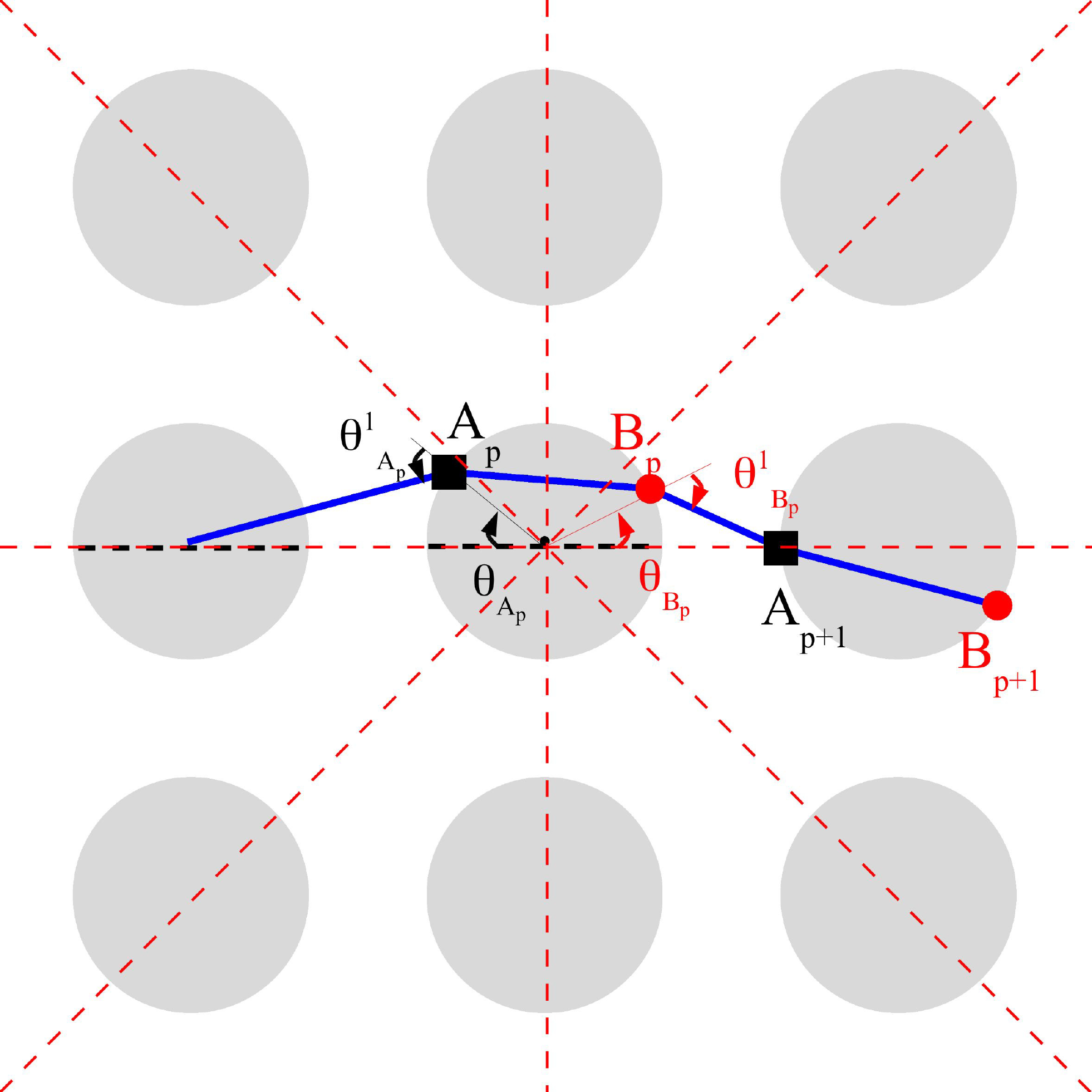}
\caption{A small portion of the photonic crystal and notations used in the text. "$A_p$" referred to the $p^{th}$ incoming point on the cylinder surface at the $p^{th}$ step, "$B_p$" is the $p^{th}$ point at the cylinder surface after refraction. $\alpha$ denotes the angle between the ray and the horizontal axis. $\theta_{A_p}$ (\textit{resp.} $\theta_{B_p}$) the angle of the point $A_p$ (\textit{resp.} $B_p$) and the horizontal axis. $\theta_{A^1_p}$ (\textit{resp.} $\theta_{B^1_p}$) the angle of the point $A_p$ (\textit{resp.} $B_p$) the angle corresponding  to the direction under which the ray intersect the cylinder. The dashed red lines are the axis of symmetry of the crystal.}
\label{Fig:schema}
\end{center}
\end{figure}

\section{The Photonic Billiard}

The photonic crystal under study is depicted in Fig. \ref{Fig:schema}: it is a biperiodic set of dielectric cylinders, in which the trajectories of a ray are studied, by applying Snell-Descartes relations at the boundary of the cylinders. If the cylinders are perfectly reflecting instead of transparent then the system is nothing else than a periodic Lorentz gas. It has an extensive bibliography both in the physical and mathematical literatures (see e.g. \cite{Gaspard} and references hereby). It consists in an ensemble of noninteracting point particles moving freely with elastic collisions on fixed scatterers (the cylinders). The photonic billiard can be seen as the refractive extension of the periodic Lorentz gas. Upon this analogy a light ray represents a particle trajectory and, instead of having infinite walls, the repulsive potential is finite with a value given by the optical index of the cylinders. An important difference with the periodic Lorentz gas is that now rays carry energy that is split between the refracted and the reflected ray. For moderate values of the cylinder optical index, the reflection coefficient in energy ($4\%$ for one single glass/air interface) is small. In such a case, energy is predominantly carried through the rays that are transmitted through the cylinder. See the Supplementary Materials (\ref{binomial}) for more details. Whereas there exists a huge literature about the periodic Lorentz gas with hard-wall scatterers, we have not been able to find references on the "refractive periodic Lorentz gas" \cite{note1}. The Lorentz gas is an unfolding of the Sinai Billiard\cite{Sinai} for which the dynamics is always chaotic,whatever the photonic crystal parameters, making impossible long-range precision concerning a single trajectory. So, despite its deterministic character, this system can only be studied from a statistical point of view.  Transport is diffusive and the square geometry is known \cite{Geisel01,Klafter} to exhibit superdiffusion processes (also called anomalous diffusion) i.e. diffusion for which the mean square displacement $<r^2>$ grows faster than linearly with time $t$ like $<r^2> \propto t ~ \ln t$ \cite{Bleher}. In the past few years, superdiffusion gained an increasing interest in optics \cite{Wiersma} with the observation of weak localization \cite{Wiersma02} in disordered samples designed to exhibit anomalous diffusion \cite{Wiersma03}. Our photonic billiard, or refractive Lorentz gas, shows a richer dynamics as compared to the Lorentz gas and the main result here is that the geometry studied in this paper also exhibits diffusion processes that can be tuned from nearly ballistic to brownian,  including anomalous diffusion. 
The letter is organized as followed. First the dynamics of the refractive Lorentz gas is discussed through Poincar\'e surfaces of section. The structure exhibits soft chaos, i.e. regular paths (also called ballistic paths), as well as chaotic motion as a consequence of its quasi-integrable dynamics\cite{Ozorio}. Above a threshold value of the normalized period $t^{\star}=T/R$  where $T$ is the crystal period and $R$ the cylinder radius, the dynamics becomes completely chaotic. Chaos strength is quantified through the computation of Lyapunov exponents. Diffusive properties are computed, showing that the refractive Lorentz gas exhibits superdiffusion due to the existence of Levy flights.

\section{The rays dynamics}

\begin{figure}[h]
\begin{center}
\includegraphics[width=8cm]{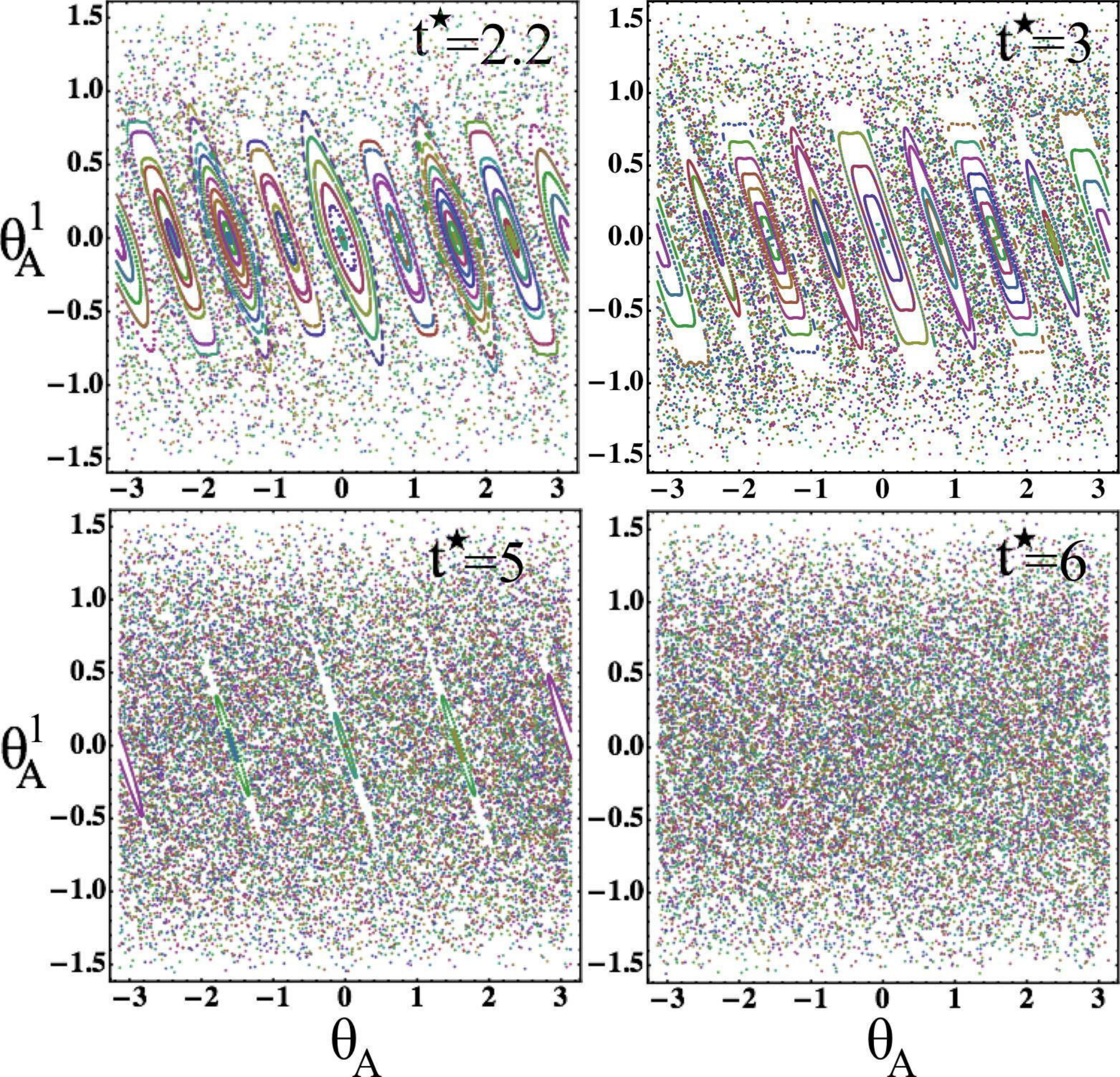}
\caption{Poincar\'e surface of section for four values of the crystal photonic period $T$ normalized by the circle radius $R$, $t^{\star}=T/R$. They result from 100 initial conditions, each evolved for 200 steps. For small normalized period (i.e. low density) the Poincar\'e surface of section are dominated by ballistic paths ($t^{\star}=2.2$) whereas the dynamics is completely chaotic for $t^{\star}$ higher than $t^{\star}=6$ here. The area of islands of stability decreases when increasing $t^{\star}$ (examples are given for $t^{\star}=3$ and $t^{\star}=5$). For $t^{\star}=5$ islands around $\theta_A=\pi/4 [\pi]$ no more exist because $t^{\star} \sqrt{2} \geq  \frac{2n}{n-1}$ as explain in the text.}
\label{Fig:Poincare}
\end{center}
\end{figure}

The dynamical properties of rays in the crystal are now described. Numerical computations were performed with an optical index of the cylinders equal to $n=1.5$, which leads to a coefficient of reflection of the order of $4~\%$ per air/glass interface. In Fig:\ref{Fig:Multi} of the supplementary materials( \ref{gaussianbeam}), we show the propagation of a laser beam computed by solving rigorously the Maxwell equations into the photonic crystal. Because of the low optical-index contrast between the cylinder and the surrounding medium, light propagation is dominated by the transmitted rays. So we focus only on rays which are refracted and thus transmitted through the cylinders. The ray dynamics has a Hamiltonian formulation in terms of the eikonal equation: $\frac{d}{ds}\left( n \frac{d \gamma}{ds}\right)=\nabla n$ \cite{Landau}, therefore the phase space is four dimensional. Because $n$ is constant inside the cylinder, the solutions to this equation, i.e. the ray trajectories, are piecewise linear. In order to characterize them, we compute the Poincar\'e surface of section. Poincar\'e surface of section are a convenient way to represent as a 2D-graph a higher dimensionality phase-space by representing a continuous evolution in a discrete way. They preserve informations about the dynamics\cite{chaosbook}. As an example, periodic orbits in the phase space appears as periodic points in the Poincar\'e surface of section\cite{Ozorio}. For the refractive Lorentz gas, the Poincar\'e surface of section is the set of points $(\theta_{A_p},\theta_{A^1_p})$ defined by the intersection of the ray with the boundary of a cylinder:  $\theta_{A_p}$ is the angle between the horizontal axis and the light ray when it crosses the boundary of the cylinder and $\theta_{A^{1}_p}$ is the angle between the incident ray and the normal to the cylinder boundary (see Fig:\ref{Fig:schema}).  
The Poincar\'e map is constructed in two steps (see the supplementary materials (\ref{map})) : 
\begin{enumerate}
\item An incident ray intersects a cylinder, defining a point $A_p$ given by the system of equations (2a and 2b),
\item The ray is refracted twice at the boundary, giving the point $B_p$. The set of equations (1a and 1b) describes these refraction steps. 
\end{enumerate}
The ray propagates in air until it hits a new cylinder, this gives $A_{p+1}$ and the process goes on. The equations are solved numerically in order to compute the ray trajectory in the whole crystal.

Examples of Poincar\'e surfaces of section are given in Fig:\ref{Fig:Poincare} for different values of the normalized period $t^{\star}$ ($t^{\star}$ varying from 2.2 to 6). It will be shown in the following that this parameter governs the stability properties of the ray dynamics.

\begin{figure}[h]
\begin{center}
\includegraphics[width=7.5cm]{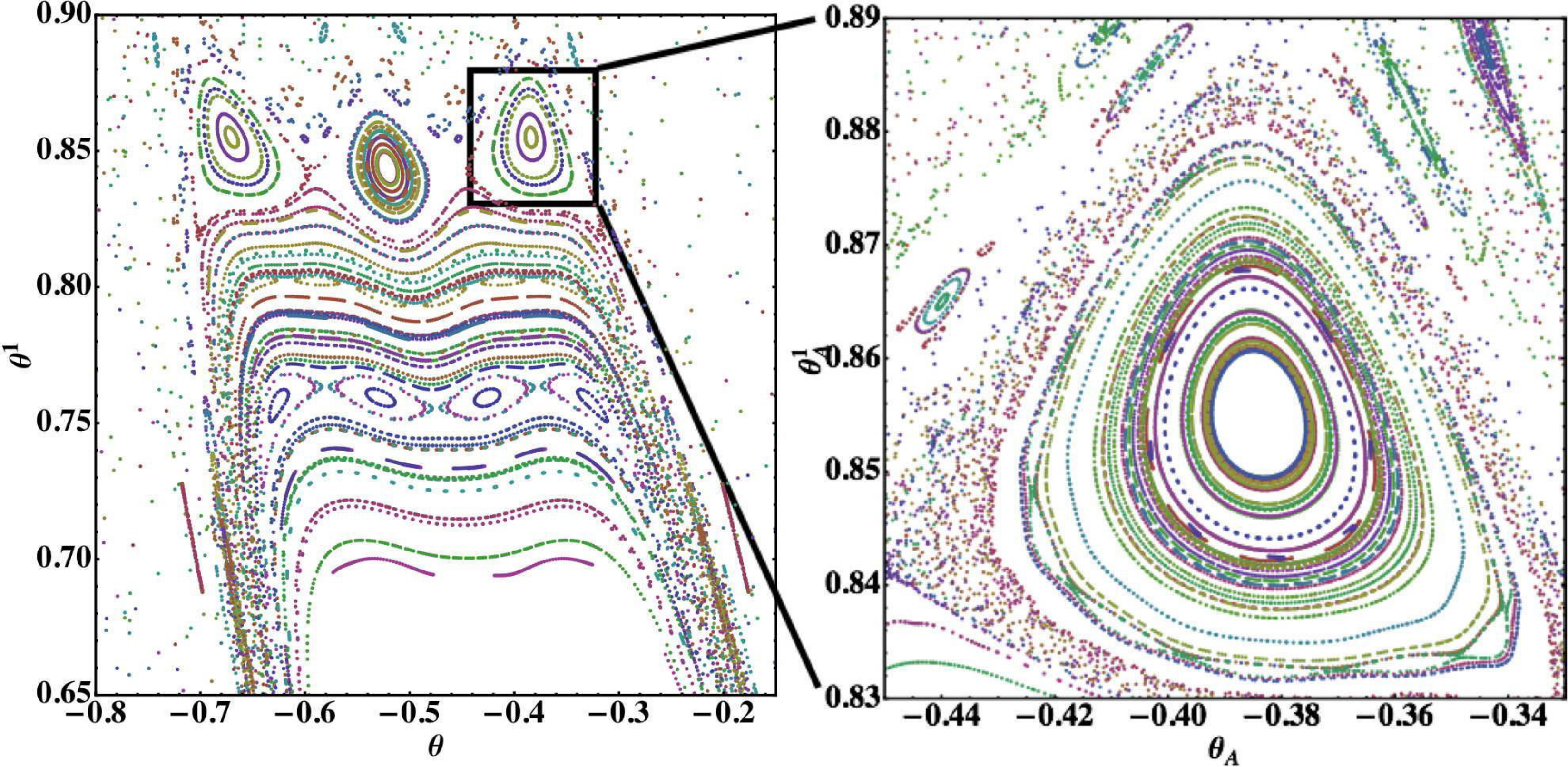}
\caption{Details of Fig:\ref{Fig:Poincare} for $t^{\star}=3$ showing Cantori surrounded by the chaotic sea.}
\label{Fig:KAM}
\end{center}
\end{figure}

The most prominent point is the existence of islands of stability characterized by closed curves in the Poincar\'e surface of section, surrounded by a chaotic sea. Islands of stability characterize ballistic trajectories in the crystal, i.e. rays that oscillate around the axis of symmetry of the crystal. Cylinders act as cylindrical lenses: they focus rays in the vicinity of the axis of symmetry of the crystal. As a result, these trajectories are periodic or quasi-periodic in real space as explained in the supplementary materials (\ref{quasi}) and illustrated by the figures Fig:\ref{Fig:period} and Fig:\ref{Fig:quasi}. The area covered by these islands of stability decreases when  $t^{\star}$ increases and they reduce to single points above the threshold $t^{\star}_{thres} \geq  \frac{2n}{n-1}$ as shown in  the supplementray materials \ref{thres}. In the example of Fig:\ref{Fig:Poincare}, the threshold is $t^{\star}_{thres}=6$ for $n=1.5$.  At small angles, the islands of stability are ellipsis. This corresponds to Gauss conditions: $\sin(\theta) \sim \theta$ and $\cos(\theta) \sim 1$. When the angles increase, the islands of stability are deformed because of non-linear effects, although rays still propagate along the crystal axis. At the edges of the transition between the islands of stability and the chaotic sea, Cantori \cite{note2} can be observed (see Fig:\ref{Fig:KAM}) because of an increase of non-linear effects: this is a direct consequence of KAM theory \cite{Ozorio}. Again rays propagate along the crystal axis. When non-linear effects still increase, rays leave crystal axis and the motion is completely chaotic, with sometimes segments of ballistic motion. In that case rays diffuse in the entire crystal. We conclude that the "quasi-integrable" character of the light dynamics in a photonic crystal \cite{Cruz} is a consequence of the dynamical properties of the transmitted rays.

\begin{figure}[h]
\begin{center}
\includegraphics[width=7cm]{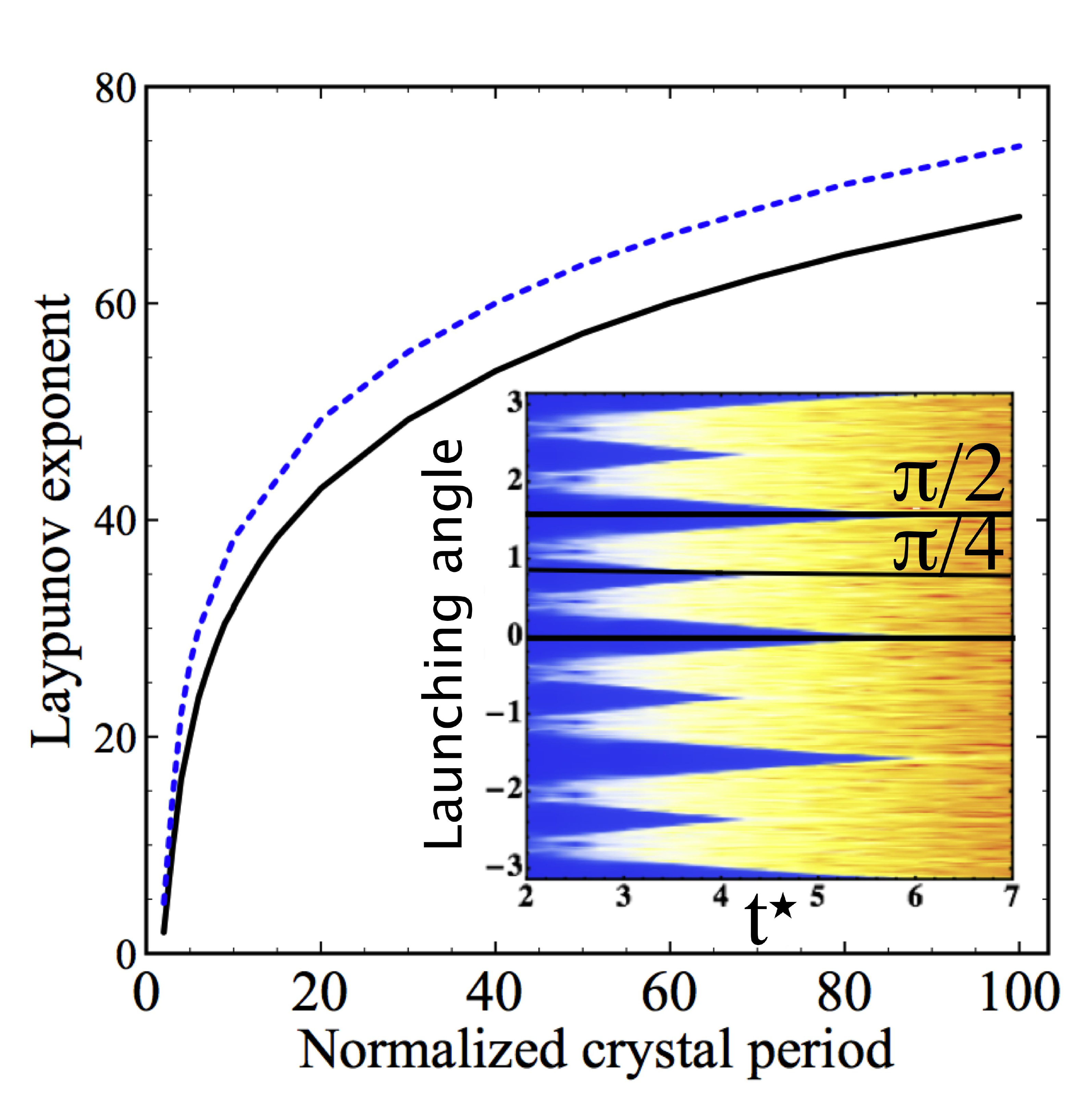}
\caption{Lyapunov exponent versus the normalized period $t^{\star}$ of the crystal. The plain line is the Lyapunov exponent for the refractive Lorentz gas whereas the dashed blue curve corresponds to the reflective Lorentz gas. Lyapunov exponents are given per step unit. The inset shows the Lyapunov exponent for the refractive Lorentz gas as a function of $t^{\star}$ and of the initial angle between the ray and the horizontal axis (launching angle). It takes low values (blue color) for launching angles around the crystal axis of symmetry. For $t^{\star} \geq t^{\star}_{thres}=6$ the Lyapunov exponent is independent of the launching angle.}
\label{Fig:Lyap}
\end{center}
\end{figure}

The strength of chaos, \textit{i.e.} the sensitivity to initial conditions, can be quantified through the computation of Lyapunov exponents. In Hamiltonian systems with two degrees of freedom, Lyapunov exponents come in pairs with a null sum because of area preservation in the phase-space \cite{Eckmann} (this is Liouville theorem). The positive exponent denoted $\lambda$, and called in the following Lyapunov exponent, quantifies the exponential sensitivity to initial conditions. The Lyapunov exponents $\lambda$ were computed per step, that is, each time a ray hits a cylinder.  Because of islands of stability, the dynamics is clearly not ergodic and the Lyapunov exponent may depend on initial conditions. The Lyapunov exponent takes low value for ballistic paths, as expected (blue color in the lower inset of Fig: \ref{Fig:Lyap}) and the value sharply increases outside the islands of stability. As shown in Fig:\ref{Fig:Poincare}, the size of the islands of stability decreases as $t^{\star}$ increases. 

Above this threshold, the Lyapunov exponent is almost constant and large, whatever the initial conditions are.  In Fig:\ref{Fig:Lyap}, the Lyapunov exponents averaged over the initial conditions are represented (plain line).  The averaged exponents  increase monotonically when increasing the normalized period $t^{\star}$. No apparent discontinuity in the behavior of the Lyapunov exponent is seen near the threshold. Its value is smaller than the Lyapunov exponent for the reflective Lorentz gas (dashed line in Fig:\ref{Fig:Lyap}) but their behavior with $t^{\star}$ is similar. As a matter of fact for large normalized period, both of them grow as $\lambda \sim  \ln(t^{\star})$ [see \cite{Gaspard} for reflective Lorentz Gas and the supplementary materials (\ref{lambaAsymp}) and the figure Fig:\ref{Fig:Asymp_Lyap} for the refractive Lorentz Gas]. This is due to the fact that for large $t^{\star}$ refractions at cylinders become uncorrelated and the distance between two refractions is given by the mean free path which is identical for both the reflective and the refractive periodic Lorentz gases. As an example, the averaged Lyapunov exponents is $\lambda=16.7$ for $t^{\star}=6$ (see Fig:\ref{Fig:Lyap}). This value means that, starting with an initial accuracy of $\Delta\theta^{1}_0=10^{-14}$ radian for the initial angle, after hitting only two cylinders the uncertainty on the angle becomes $\Delta \theta^{1}_2=10^{-14} \times e^{2 \times 16.7} = 3.2$. Thus, after hitting only two cylinders, the value of the angle $\theta^1$ could be any value in $[-\pi,\pi]$. We have performed experiments as shown in the supplementary materials (\ref{Exp}). They clearly demonstrate this exponential sensitivity to the initial conditions that can lead to the diffusion of the laser beam into the photonic crystal.
 
\section{Diffusive properties of the photonic billiard}

\begin{figure}[h]
\begin{center}
\includegraphics[width=8cm]{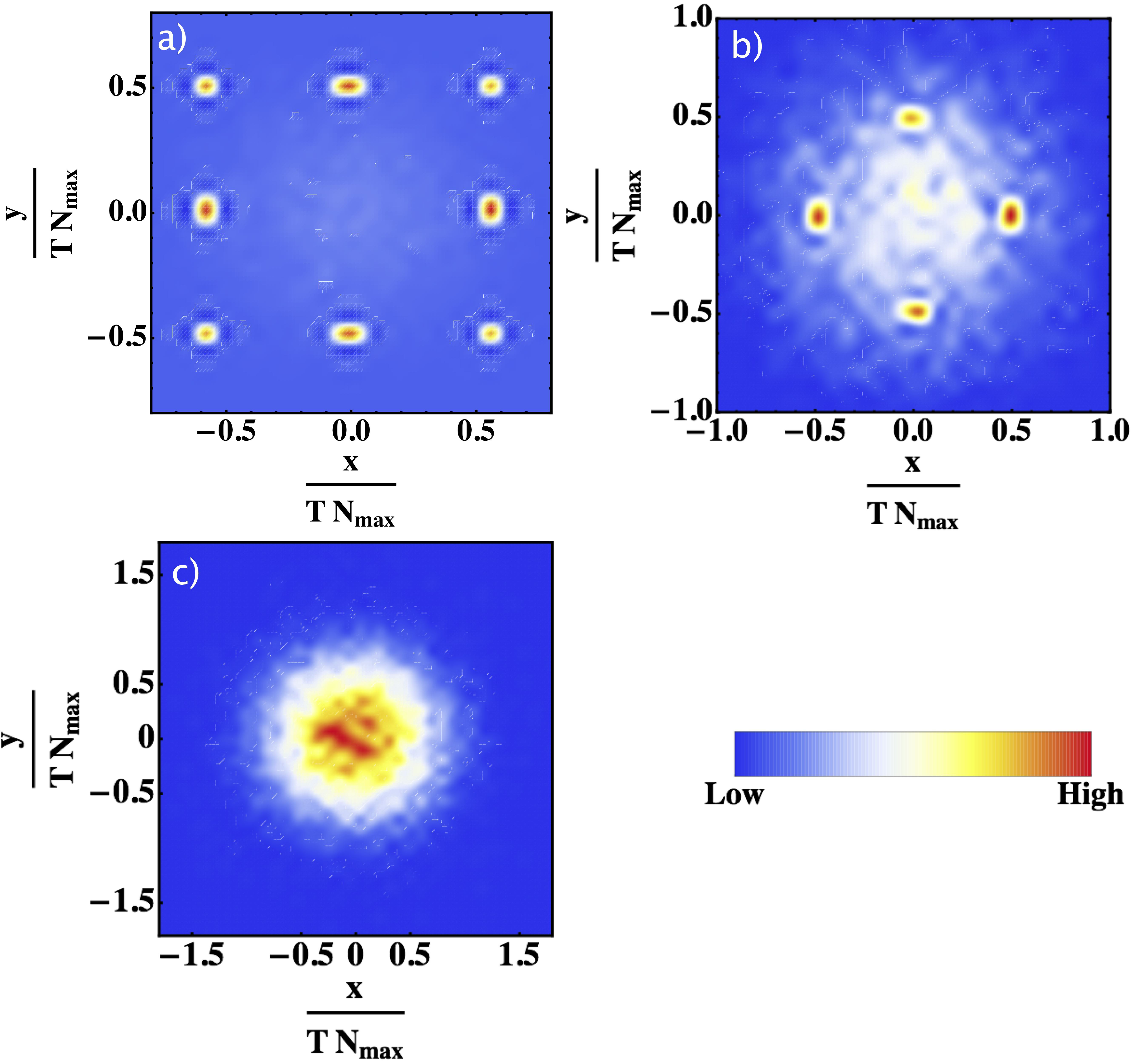}  
\caption{ Rays density after 300 simulation steps for a) $t^{\star}=3$, b) $t^{\star}=5$ and c) $t^{\star}=6$. The simulation was running with $3.10^4$ rays starting from (0,0) with random initial conditions under a uniform distribution.\label{Fig:distribution}
}
\end{center}
\end{figure}

From an experimental point of view, a light beam can be seen as an ensemble of rays with a spatial extension. The rays constituting the light beam have different directions of propagation. As a consequence, from an experimental point of view, an incoming light beam can be seen as an ensemble of rays with different initial conditions $\{ \theta_A(0), \theta_A^1(0) \}$. Because of the high values of the Lyapunov exponent, a light beam will undergo a diffusive propagation inside the photonic crystal. In order to characterize the diffusive properties of light in the photonic crystal, the propagation of a bundle of rays is computed. This simulation models the emission of a point source (\textit{e.g.} a molecule or a quantum dot) inside the photonics crystal. As we will explain in the following our numerical results also provide informations about the propagation of a light beam inside the photonic crystal through the computation of the exponent of the diffusion law. In the following, we first show how rays are distributed into the photonic crystal by computing the density of rays. Then we compute the time evolution of the mean-square displacement. The simulation is performed as followed. The rays are initially launched at the center of the photonic crystal with random initial conditions (under a uniform distribution). After $N_{max}$ steps (i.e. each ray hints $N_{max}$ times a cylinder), the density of rays (i.e. the number of rays per unit area) is plotted. For low values of $t^{\star}$ (see Fig: \ref{Fig:distribution}-a) for $t^{\star}=3$) the dynamics is barely chaotic with stability islands covering a large part of the Poincar\'e surface of section (Fig: \ref{Fig:Poincare}) and characterized by a low-value of the Lyapunov exponent (Fig: \ref{Fig:Lyap}). In that case  one can clearly notice a non-isotropic density of rays. Most of the rays concentrate in 8 spots localized on the principal axis of the photonics crystal. These rays propagate in the photonic crystal through regular trajectories. 
When $t^{\star}$ is above the threshold $t^{\star}\ge t^{\star}_{thres}$, the dynamics only admits chaotic motion (see Fig:\ref{Fig:distribution}-c) with $t^{\star}=6$) and the density of rays becomes isotropic with a maximum at the center of the photonic crystal.  The density of rays follows a Gaussian distribution law, which indicates that rays undergo a brownian motion. For intermediate values of the normalized period $t^{\star}$, the density of rays is characterized by spots on the crystal axis plus an isotropic distribution which is maximum at the photonic crystal center. For this intermediate situation, some rays follow regular paths which lead to the spots in the rays density and some others follow chaotic motion giving the isotropic distribution.

\begin{figure}[h]
\begin{center}
\includegraphics[width=7.5cm]{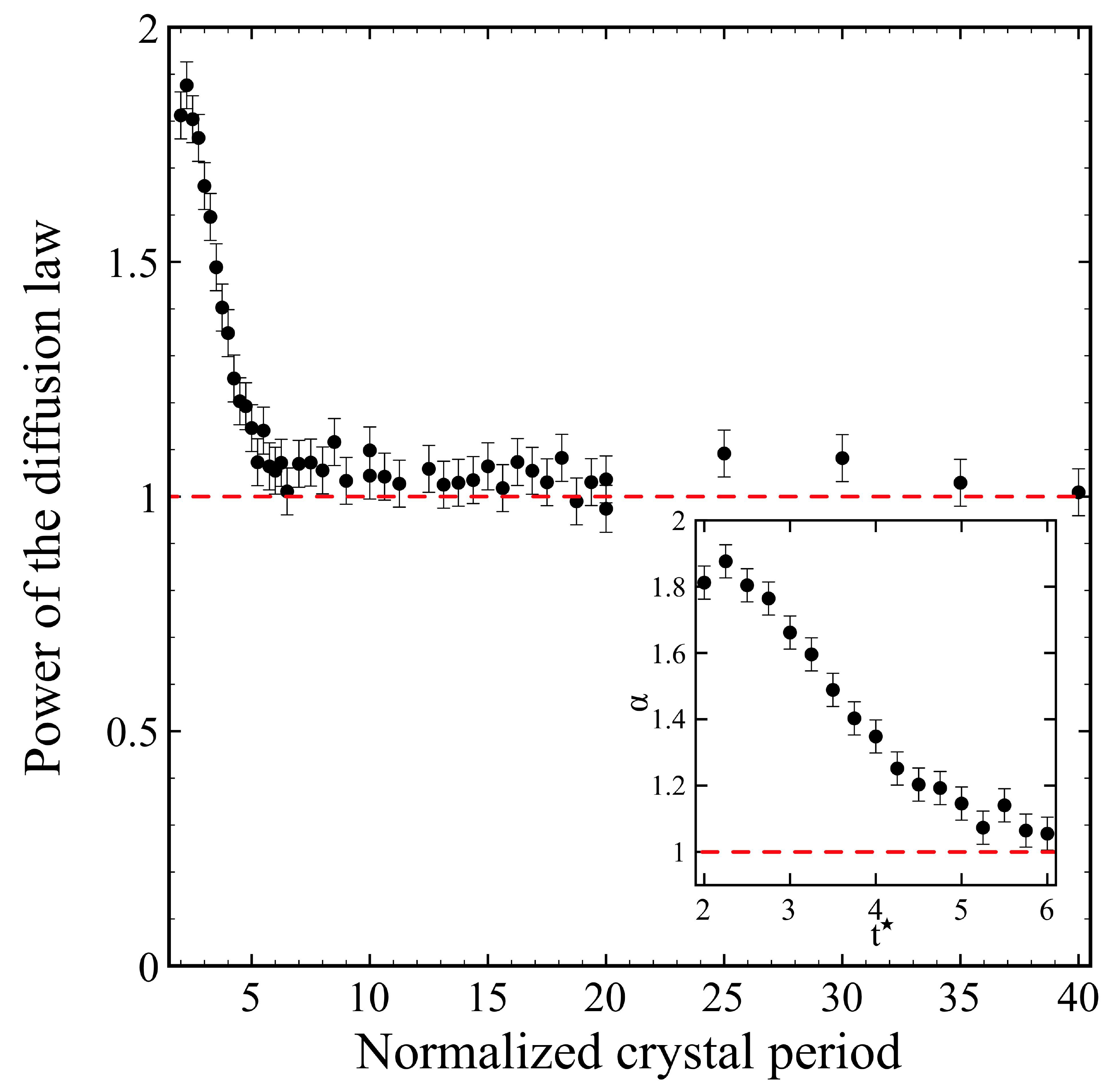}
\caption{Exponent of the diffusion law defined by the law $<(r-r_0)^2>=D_{\alpha} t^{\alpha}$ versus the normalized period $t^{\star}$ of the crystal. The inset is a zoom on the interesting region where islands of stability exist $t^{\star} \in [2,6]$.$10^{4}$ rays was used to compute the exponent of the diffusion law. }
\label{Fig:PowerLaw}
\end{center}
\end{figure}

We now describe the computation of the evolution of the mean-square displacement with time. Hamiltonian systems are known to lead to superdiffusion \cite{Geisel01}. It has been shown that in a two-dimensional periodic potential \cite{Geisel01,Klafter} the mean square-displacement varies as $<\textbf{r}_n^2> \sim t^{\alpha}$ as a results of ballistic flights, random motion and the "sticky" barriers formed by Cantori \cite{Denisov}. From an experimental point of view, let us consider a layer of width $L$ of a diffusive material. The transmission coefficient ,\textit{i.e.} the fraction of light that is transmitted through the layer of material, can be related to the exponent of the diffusion law \cite{Wiersma}. Indeed, it has been shown that the transmission coefficient is of the form $T=\frac{1}{1+A L^{(3-\alpha)/2}}$ where L is the thickness of the layer of diffusive material, $\alpha$ the exponent of the diffusion law and $A$ a constant. As a consequence, computing the diffusion law in the photonic crystal gives informations on the repartition of the light emitted by a point-source but also to the transmission of a light beam through a finite size photonic crystal. In order to find the exponent of the diffusion law $\alpha$, we compute the mean square-displacement for an ensemble of $10^4$ rays with initial conditions chosen randomly (with a uniform distribution). The simulation time is long enough to reach the stationary regime, following the prescription of ref.\cite{Sanders}. The exponent of the diffusion law is plotted on Fig:\ref{Fig:PowerLaw} as a function of the normalized period $t^{\star}$. Its values are larger than 1 when the dynamics contains chaotic regions and islands of stability, which confirms the superdiffusive behavior. It reaches values close to 2 when the gap between the cylinders is small ($t^{\star} \sim 2 $), i.e. when the dynamics is dominated by islands of stability and it decreases to 1 as the area of stability islands decreases.  For $t^{\star}\geq 6$, that is to say when the dynamics is completely chaotic, diffusion is close to normal. Nevertheless, the exponent remains slightly larger than 1 because of ballistic paths exactly along the crystal principal axis and because of infinite-horizon trajectories (\textit{i.e.} light rays that can move arbitrarily far without crossing a cylinder). Our results are in strong contrast with the periodic Lorentz gas. Indeed, the periodic Lorentz gas exhibits a weak form of superdiffusion \cite{Sanders} with a logarithmic correction to the normal diffusion\cite{Bleher, Sanders}. Moreover the diffusion exponent does not depend on the period of the periodic Lorentz gas.

\section{Conclusion}

In the high frequency limit, light propagation in a photonic crystal exhibits complex propagation patterns with regular paths and chaotic motion. For this limiting case, light transport is best described as a diffusive process that can be tuned from a superdiffusive regime to a nearly-brownian motion. An interesting feature of the system is its simplicity, allowing for experimental investigations of wave chaos footprint in optics.



\begin{thebibliography}{0}
\bibitem{note1} We use the terminology ``photonic billiard'' or ``refractive Lorentz gas'' as synonyms throughout the letter.
\end{thebibliography}
\end{document}